\documentclass[%
,twocolumn%
,secnumarabic%
,amssymb,aps,prb,
]{revtex4-1}
\usepackage{docs}

\usepackage[english]{babel}
\usepackage[utf8,latin1]{inputenc}
\usepackage{graphicx}
\usepackage{graphics}

\begin{document}
\title{Effects of rotation in the energy spectrum of $C_{60}$}
\author{Jonas R. F. Lima$^{1}$, J\'ulio Brand\~ao $^1$, M. M. Cunha $^2$ and F. Moraes$^1$}%
\affiliation{$^1$Departamento de F\'{\i}sica, CCEN,  Universidade Federal 
da Para\'{\i}ba, Caixa Postal 5008,  58051-900 , Jo\~ao Pessoa, PB, Brazil }
\affiliation{$^2$Departamento de F\'{\i}sica, CCEN,  Universidade Federal 
de Pernambuco, 50670-901 , Recife, PE, Brazil}

\date{\today}%


\begin{abstract}
In this paper, motivated by the experimental evidence of rapidly rotating $C_{60}$ molecules in fullerite, we study the low-energy electronic states of  rotating fullerene within a continuum model. In this model, the low-energy spectrum is obtained from an effective Dirac equation including non-Abelian gauge fields that simulate  the pentagonal rings of the molecule. Rotation is incorporated into the model by solving the effective Dirac equation in the rotating referential frame. The exact analytical solution for the eigenfunctions and energy spectrum is obtained, yielding the previously known static results in the no rotation limit. Due to the coupling between rotation and total angular momentum, that appears naturally in the rotating frame, the zero modes of static $C_{60}$ are shifted and also suffer a Zeeman splitting whithout the presence of a magnetic field. 
\end{abstract}
\maketitle

\section{Introduction}

$C_{60}$ molecules rotate. Even in crystalline fullerite \cite{Yannoni,Neumann}, or in ``peapods''inside single walled carbon nanotubes \cite{Zou}, or in molecular layers on top of crystals
\cite{Sanchez}, 
they present quasi-free rotation even down to low temperatures. The orientational disorder of the rotating $C_{60}$ molecules  made the crystalline structural determination of  the solid state be completed  only about  five years after their discovery. This was achieved by smartly breaking the spherical symmetry of $C_{60}$ by the addition of a functional unit yielding as result  a crystal with orientational order \cite{HAWKINS,Fagan}.

Solid fullerene, or fullerite $C_{60}$,  is a very unusual solid in the sense that it is made of rotationally mobile molecules. Below 249K it has an orientationally ordered phase reminiscent of a ferromagnet or a nematic liquid crystal. Near $T_c=250-260K$ it presents a first order transition to the orientationally disordered phase similar to a paramagnet or the liquid crystalline isotropic phase.  At room temperature, the face-centered cubic (fcc) structure of fullerite has, in each cubic unit cell, four molecules rotating nearly freely at frequencies about \cite{JOHNSON} $10^{11}$ Hz. Much new physics comes from this rotational degree of freedom. The rapid rotation has important consequences on the nanotribological properties of fullerite  single crystal surfaces \cite{Liang}, on the interaction between the $C_{60}$ molecules \cite{Yan} and on the electronic transport in the crystal \cite{Katz}. 

\begin{figure}[hpt]
\centering
\includegraphics[width=6cm,height=8cm]{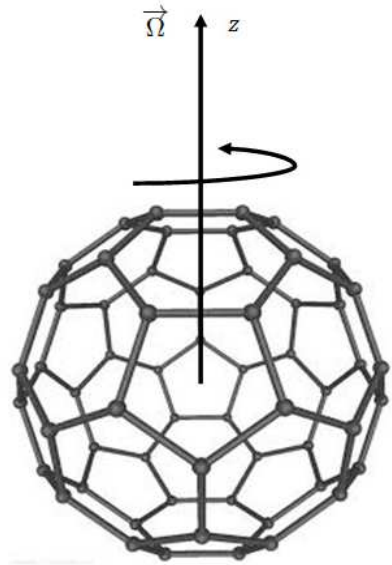}	
\caption{Rotating Fulerene. The angular velocity $\vec{\Omega}$ is in the $z$ direction.}\label{fulereno}
\end{figure}

Even though rotation is so important to the determination of the physical characteristics of $C_{60}$, very few works \cite{shen,shen2} have been published on its influence on electronic properties of the molecule. An obvious question is how rotation affects the electronic energy spectrum of $C_{60}$. This question was addressed by Shen and Zhang \cite{shen} who computed, by aproximative methods, an energy shift proportional to the rotational angular velocity of the molecule. 

In this article, using the continuum approach for the electronic structure of $C_{60}$, we confirm the result of Shen and Zhang and show that rotation also introduces a Zeeman-like spliting of the energy levels. As a consequence, the zero modes predicted \cite{vozmediano,vozmedianoprl,osipov} for static $C_{60}$, no longer appear. Nevertheless, in the zero rotational angular velocity limit, the zero modes, as well as the stactic $C_{60}$ energy spectrum are recovered.  We use a Dirac equation continuum model based in the tight-binding approximation, first presented in \cite{vozmedianoprl,vozmediano}, which is limited to electronic states close to the Fermi level. In this model the lattice structure disappears and the pentagons in $C_{60}$ are simulated by localized fictitious gauge fluxes. 

The organization of the paper is as follows. In Section II we describe the continuum model for rotating fullerene, obtaining the Dirac operator for the problem. The solution of the eigenvalue problem is presented in Section III, where we find an exact analytical solution for both, the eigenfunction and the energy spectrum. Finally, in Section IV we present our conclusions.

\section{The continuum model for  rotating fullerene}

\subsection{Graphene: from tight-binding to field theory}

Our starting point is the fact that, in the long wavelenght limit,  the electronic spectrum of  graphene can be obtained from an effective Dirac equation in (2+1) dimensions. This was found using the tight binding \cite{wallace} and effective-mass \cite{DiVincenzo} approximation. From the tight-binding Hamiltonian for electrons in graphene, considering only nearest-neighbour hopping,  the energy bands are given by \cite{RevModPhys.81.109}
\begin{eqnarray}
E_\pm=\pm &t&\Biggl(3+2\cos(\sqrt{3}k_ya)   \nonumber  \\  
 &+&4\left. \cos(\frac{\sqrt{3}}{2}k_ya)\cos(\frac{3}{2}k_xa)\right)^{1/2},
\label{equation}
\end{eqnarray}
where $t$ is the nearest-neighbour hopping energy and $a$ is the carbon-carbon distance. Expanding this relation around the Dirac points $\vec{K}$  in the Brillouin zone, {\it i.e.}, writing $\vec{k}=\vec{K}+\vec{q}$, with $\mid\vec{q}\mid\ll\mid\vec{K}\mid$, one obtains
\begin{equation}
E_\pm(\vec{q})=\pm v_F\mid\vec{q}\mid+O[(q/K)^2].
\label{dispersion}
\end{equation} 

This linear dispersion relation justifies the effective theory  for the low-energy electrons  of graphene as two-dimensional massless Dirac fermions. The electronic states are represented by the two-component wave function $\psi=(\psi_A,  \psi_B)^T$, called \textit{pseudospin}, which represents the two graphene sublattices $A$ and $B$. The wave function $\psi$   obeys the massless Dirac equation  in two dimensions
\begin{equation}
-i \hbar v_F\sigma^\mu\partial_\mu\psi(r)=E\psi(r),
\label{dirac1}
\end{equation}
where $\sigma^\mu$ are the Pauli matrices ($\mu=1, 2$) and $v_F$ is the Fermi velocity.  

Besides  the pseudospin, there is a second spin-like entity due to the fact that there are two independent wave vectors, $\vec{K}_+=\vec{K}$ and  $\vec{K}_-=-\vec{K}$, which give the same dispersion relation (\ref{dispersion}). This degree of freedom is called $K$-spin. The tensor product between the two ``spin'' spaces yields a four dimensional space expanded by the kets $\vert\vec{K}_\pm A\rangle$ and $\vert\vec{K}_\pm B\rangle$. Therefore, each wave function $\psi_A$ and $\psi_B$ will have two subcomponents related with the two Dirac points: $\psi_A=(\psi_{A}^{K_+}, \psi_{A}^{K_-})^T$ and $\psi_B=(\psi_{B}^{K_+}, \psi_{B}^{K_-})^T$. The Pauli matrices which act on the $K$ part of the wave function will be denoted by $\tau^\mu$.  The $K$-spin plays an important role in one of the fictitious gauge fields flux related to the existence of pentagons in the fullerene molecule. This will be seen in the next subsections where the Dirac operator (\ref{dirac1}) will be modified in order to incorporate the spherical geometry,  the disclinations (pentagons)  and  rotation of the fullerene molecule.

\subsection{Disclinations: from graphene to fullerene}
 
 \begin{figure}[hpt]
\centering
\includegraphics[width=5.5cm,height=5cm]{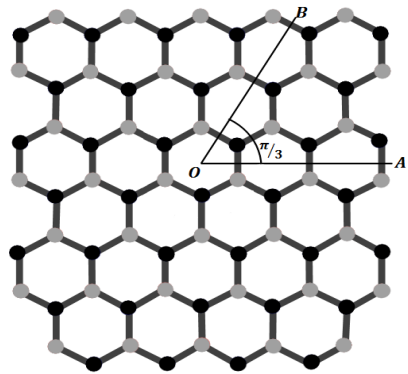}	
\caption{Cutting a $\pi/3$ angular section of the graphene sheet. Connecting the dangling bonds on the sides OA and OB one gets a cone with a pentagon on the vertex. Grey and black circles represent sublattices A and B, respectively.}\label{corte}
\end{figure} 

To transform a graphene sheet into a $C_{60}$ fullerene one needs to introduce curvature into the  hexagon-tiled plane. If we cut a $\pi/3$ wedge out of the plane and join the dangling bonds at the opposing edges (see Fig. 2), a cone is created with a pentagon at its tip. The pentagons are then defects, called disclinations, in the otherwise perfect hexagonal tiling of the graphene sheet. In the continuum limit, each disclination carries a $\delta$-function curvature singularity \cite{vickers}
\begin{equation}
R_{12}^{\,\,\,\,\,\,\,12}=\lambda \delta (x)\delta(y), \label{R12}
\end{equation}
where $\lambda$ is the angle characterising the removed wedge which, in the case of the pentagon, is $\frac{\pi}{3}$.

To get a spherical shape, 12 pentagons, symmetrically arranged are needed. The result is $C_{60}$ which, in the continuum limit, has 12 curvature singularities which integrate out to give the total curvature of the sphere.  These curvature singularities act as flux tubes giving rise to a Aharonov-Bohm-like phase\cite{carvalho,osipov1992}.  Therefore, the ``elastic"  geometric  fluxes incorporate the topological origin of the defects and should be included in the Dirac equation (\ref{dirac1}). 

Disclinations are characterized by the Frank vector $\Theta_i$. In the geometric theory of defects \cite{Katanaev} the curvature associated to the disclination is the surface density of Frank vectors, such that
\begin{equation}
\Omega^ {ij}=\int\int dx^{\mu} \wedge dx^{\nu} R_{\mu\nu}^{\,\,\,\,\,\,ij}, \label{Omega}
\end{equation}
with 

\begin{equation}
\Theta_i = \epsilon_{ijk}\Omega^ {jk}. \label{Thetai}
\end{equation}
The curvature tensor, in terms of the $SO(3)$ connection $\omega_{\nu j}^{\,\,\,\,\,\,i}$,  is 
\begin{equation}
R_{\mu\nu j}^{\,\,\,\,\,\,\,\,\,i}=\partial_{\mu}\omega_{\nu j}^{\,\,\,\,\,\,i} - \omega_{\mu j}^{\,\,\,\,\,\,k}\omega_{\nu k}^{\,\,\,\,\,\,i} - (\mu \leftrightarrow \nu). \label{Rtensor}
\end{equation}
Since we have a two-dimensional system, the rotations are restricted by the subgroup $SO(2)$, which is Abelian. This  implies that the quadratic terms in (\ref{Rtensor}) vanish and the curvature tensor can be seen as the curl of the $SO(2)$ connection. This way, the Frank vector is then given by
\begin{equation}
\Omega^ {ij}= \oint dx^{\mu} \omega_{\mu}^{\,\,\,\,\,ij}. \label{omegaij}
\end{equation}
Using (\ref{R12}), (\ref{Omega}), (\ref{Thetai}) and $\lambda=\frac{\pi}{3}$ one gets
\begin{equation}
\Theta^3 = \frac{\pi}{3}.
\end{equation}
Equivalently, in the gauge theory of disclinations  \cite{fieldW} the Frank vector is obtained from  the flux of an Abelian gauge field or, the circuit integral of its  vector potential $W_{\mu}$, 
\begin{equation}
\Theta^3 = \oint dx^\mu W_\mu = \frac{\pi}{3}, \label{flux1}
\end{equation}
analogous to (\ref{omegaij}), which gives a geometric interpretation to the gauge field $W_{\mu}$. From here on we proceed like \cite{osipov} including the Abelian gauge field $W_{\mu}$ in the Dirac equation via minimal coupling.

A second and third gauge fields are needed to fix the jump in the wave function on the cone as one goes around its tip. As one can see in Fig. \ref{corte}, the alternating black and gray sublattices of graphene will have a discontinuity after the wedge is removed and the dangling bonds are joined. Gray atoms will be connected to gray atoms at the junction of the bonds. The pseudospin and K-spin parts of the wave function acquire  phases when passing the gray-gray junction as the cone tip is circulated. In order to make the discontinuity disappear in the continuous model these phases need to be removed. This is done with the help of  fluxes through the apex of the cone of fictitious non-Abelian gauge fields $\omega_{\mu}$ and $a_{\mu}$, designed \cite{Lammert} to compensate those phases. The fluxes produced by these fields are
\begin{equation}
\oint dx^\mu \omega_\mu  = -\frac{\pi}{6}\sigma^3     \label{flux2}
\end{equation}
and
\begin{equation}
 \oint dx^\mu a_\mu  = \frac{\pi}{2}\tau^2 , \label{flux3}
\end{equation}
where the integrals should be computed along closed curves around the vertex $O$ in Fig. \ref{corte}. Notice that $\sigma^3$ is the usual Pauli matrix acting on the pseudospin and $\tau^2$ is also the Pauli matrix, this time acting on the $K$-spin. The fluxes (\ref{flux1}), (\ref{flux2}) and (\ref{flux3}) modify the wave function as the Aharonov-Bohm flux $\oint dx^{\mu} A_{\mu}$ does, as generators of infinitesimal rotations in the respective Hilbert spaces where they act. 

The $\omega_{\mu}$ field is purely geometrical. In fact, it is  the spin connection, which is part of the  the covariant derivative in curvilinear coordinates. So it is naturally included in the Dirac equation on the sphere and therefore it does not contribute to the total flux.  Each of the twelve disclinations of $C_{60}$ contributes with a flux of each kind, (\ref{flux1}) and (\ref{flux3}). The problem of incorporating these discreet fluxes into the Dirac equation may be approximated by replacing them altogether by their average over the sphere. That is, effectively substituting them with the flux of a ('t Hooft-Polyakov) monopole at the center of the sphere \cite{osipov,vozmedianoprl,vozmediano}.  The electronic structure of $C_{60}$ near the Fermi level may then be obtained by solving Dirac equation on the sphere including the  gauge field of the monopole.

Like the electric charge, which is $\frac{1}{4\pi}$ times the flux of electric field through a closed surface enclosing it, the monopole charge can be obtained from the expressions (\ref{flux1}) and (\ref{flux3}), respectively:
\begin{equation}
g_W=\frac{1}{4  \pi}\cdot12 \cdot \frac{\pi}{3}=1
\end{equation}
for the  $W_{\mu}$ field and
\begin{equation}
g_a=\frac{1}{4\pi} \cdot12 \cdot \frac{\pi}{2}=\frac{3}{2}
\end{equation}
for the $a_{\mu}$ field.

In spherical coordinates, the Abelian gauge field $W_{\mu}$ is  given by \cite{osipov}
\begin{equation}
W_\theta = 0 , W_\varphi = g_W \cos\theta \label{W}
\end{equation}
and the non-Abelian gauge field $a_{\mu}$ by \cite{osipov,vozmedianoprl,vozmediano}
\begin{equation}
a_\theta = 0 , \\ \\ \\ a_\varphi = g_a \cos\theta \tau^2 . \label{a}
\end{equation}

The free particle Dirac operator on the sphere is \cite{abrikosov} 
\begin{equation}
-i\hbar c \left[ \sigma_x \left( \partial_{\theta} + \frac{\cot \theta}{2} \right) + \frac{i\sigma_y}{\sin \theta} \partial_{\varphi} \right].
\end{equation}
Therefore, the equivalent of Eq. (\ref{dirac1}) for fullerene is  \cite{osipov}
\begin{eqnarray}
-i \hbar v_F    \Bigl[   &\sigma_x &\left( \partial_{\theta}  + \frac{\cot \theta}{2} \right)  \nonumber \\ 
&+&\left. \frac{\sigma_y}{\sin \theta} \left(\partial_{\varphi}-ia_{\varphi}-iW_{\varphi}\right) \right] \psi =E\psi .
\label{equation2}
\end{eqnarray}

\subsection{Rotation}

Since we want to study the effects of rotation on the electronic spectrum of $C_{60}$ we need to solve Dirac equation in the rotating reference frame attached to the molecule. In order to obtain some insight into the not so intuitive motion in a non-inertial frame, we start this subsection by reviewing the classical physics of a free particle in a rotating medium. 

The  velocity transformation between the static laboratory frame (primed) and the rotating frame (unprimed) is 
\begin{equation}
\vec{v}\,' =\vec{v} + \left( \vec{\Omega} \times \vec{r} \right).
\end{equation}
That is, the speed of the particle as measured in the laboratory is the sum of its speed in the rotating frame plus the speed $\vec{\Omega} \times \vec{r}$ of the motion done with the rotating frame. We consider that the origin of both frames coincide with the center of the molecule. This leads to the following Lagrangian for the free particle in the rotating frame \cite{landau2}
\begin{equation}
L=\frac{mv^2}{2}+ m\vec{v}\cdot (\vec{\Omega}\times\vec{r})+ \frac{m}{2}(\vec{\Omega}\times\vec{r})^2 . \label{lagr}
\end{equation}
This gives the canonical momentum
\begin{equation}
\vec{p}=\frac{\partial L}{\partial \vec{v}}=m\vec{v}+m(\vec{\Omega}\times\vec{r}) .\label{cmomentum}
\end{equation}
Using (\ref{lagr}) and (\ref{cmomentum}) one obtains the classical Hamiltonian for the free particle in the rotating frame
\begin{equation}
H=\vec{p} \cdot \vec{v} - L = \frac{mv^2}{2}-\frac{1}{2}m(\vec{\Omega}\times\vec{r})^2 , 
\end{equation} 
or, in terms of $\vec{p}$,
\begin{equation}
H=\frac{p^2}{2m}-\vec{\Omega}\cdot\vec{L} , \label{hamil2}
\end{equation}
where $\vec{L}=\vec{r}\times\vec{p}$ is the orbital angular momentum of the particle. Equation (\ref{hamil2}) expresses also the form of the Schr\"odinger Hamiltonian for the free particle in the rotating frame, as seen in reference \cite{ni} which also gives the Dirac Hamiltonian
\begin{equation}
H = \beta mc^2+ c \vec{\alpha} \cdot \vec{p} - \vec{\Omega}\cdot (\vec{L}+\vec{S}) , \label{diracH}
\end{equation}
where $\vec{S}$ is the real spin operator and $\vec{\alpha} $ and $\beta$ are the Dirac matrices. This is essentially  the free particle Dirac Hamiltonian plus a coupling between rotation and total angular momemtum, like in  (\ref{hamil2}).

Considering the rotation axis to be in the $z$ direction we then have, from (\ref{equation2}) and (\ref{diracH}), that for the rotating fullerene
\begin{eqnarray}
& -  & i \hbar v_F    \left[ \sigma_x \left( \partial_{\theta} + \frac{\cot \theta}{2} \right) \right. \nonumber \\
&  + & \left. \frac{i\sigma_y}{\sin \theta} \left(\partial_{\varphi}-ia_{\varphi}-iW_{\varphi}\right) 
   \right] \psi -   \Omega J_z  \psi =E\psi ,
\label{equation}
\end{eqnarray}
where
\begin{equation}
J_z   =  -i\hbar \partial_{\varphi}  + S_z \label{jz}
\end{equation}
is the $z$-component of the total angular momentum. The innocent looking expression (\ref{jz}) is not as obvious as it seems. The orbital angular momentum is obtained from the mechanical moment $\vec{\pi}$  as $\vec{L}=\vec{r}\times \vec{\pi}$. For a  particle of charge $q$ moving in the presence of an ordinary magnetic field, $\vec{\pi}=\vec{p}-\frac{q}{c}\vec{A}$, where $\vec{A}$ is the vector potential associated to the magnetic field. For a particle moving in the presence of a magnetic monopole field of  charge $g$, the operator $\vec{L}=\vec{r}\times \vec{\pi}$ is no longer the angular momentum since it does not obey the usual commutation relations of the angular momentum, which must be respected in order to guarantee its conservation. It is then necessary to fix this by adding to it the angular momentum stored in the magnetic field of the monopole \cite{jackson}, $-g\hbar \hat{r}$,  so as to fix this problem. In the same way, non-Abelian monopoles contribute to the mechanical momentum via minimal coupling and also store angular momentum in their fields \cite{goddard}. Taking this into consideration, we have \cite{vozmediano,osipov}
\begin{eqnarray}
J_z    = &  - & i\hbar \left(\partial_{\varphi}-\frac{1}{4}\left[ \sigma_x , \sigma_y \right] cos\theta -ia_{\varphi}-iW_{\varphi}\right)\nonumber \\
& + & \frac{\hbar}{2} \sigma_z \cos \theta +  \hbar g_a \tau^2 \cos\theta +\hbar g_W \cos\theta + S_z
\end{eqnarray}
Taking into consideration (\ref{W}) and (\ref{a}) and the fact that $\left[ \sigma_x , \sigma_y \right] =2i\sigma_z $, one gets (\ref{jz}).

\section{Electronic Structure}

In equation (\ref{equation}) the only operator  acting in the $K$ part of the wave function is $\tau^2$ which appears only in the gauge field $a_{\varphi}$. So, the equation is already  diagonalized for this operator and we can replace it with its eigenvalue $k=\pm 1$, which correspond, respectively, to the $K$-spin states $\vec{K}_+$ and $\vec{K}_-$. In the same way  the real spin operator $S_z$ appears only in the total angular momentum $J_z$ and is replaced with the eigenvalue $s_z= \frac{\hbar}{2}s$, where $s=+1(-1)$ for spin up(down). The result is the two-component Dirac equation 
\begin{eqnarray}
& - & i \hbar v_F\sigma_{x}\left(\partial_{\theta}  +  \frac{\cot\theta}{2}\right) \psi  -   \frac{i \hbar v_F\sigma_{y}}{\sin\theta}(\partial_{\varphi} - i A \cos\theta) \psi \nonumber \\
& + & i \hbar \Omega \partial_\varphi  \psi - \frac{s \hbar \Omega}{2} \psi = E \psi
\label{dirac2}
\end{eqnarray}
where  $A=(a^k_\varphi+W_\varphi)/\cos\theta=\frac{3}{2}k+1$ with $a^k_{\varphi}=\frac{3}{2}k\cos\theta$ being the $K$-spin eigenvalue of $a_{\varphi}$. Writing
\begin{equation}
\left(
\begin{array}{c}
\psi_A \\
\psi_B \\
\end{array}
\right) = \displaystyle\sum_{j}\frac{e^{i \ell \varphi}}{\sqrt{2\pi}}
\left(
\begin{array}{c}
u_{\ell} (\theta) \\
v_{\ell} (\theta) \\
\end{array}
\right), \ell =0, \pm1, \pm2, ...
\end{equation}
we will have two coupled equations for $u_{\ell}$ and $v_{\ell}$ given by
\begin{eqnarray}
-i \hbar v_F\left(\partial_\theta + \left[\frac{1}{2}-A\right]\cot \theta+\frac{\ell}{\sin \theta}\right)v_{\ell}= \nonumber \\
\left(E+\ell \hbar \Omega +\frac{s \hbar \Omega}{2}\right)u_{\ell}
\label{acoplada}
\end{eqnarray}
and
\begin{eqnarray}
-i \hbar v_F\left(\partial_\theta + \left[\frac{1}{2}+A\right] \cot \theta - \frac{\ell}{\sin \theta}\right) u_{\ell} = \nonumber \\
\left(E+\ell \hbar \Omega + \frac{s \hbar \Omega}{2} \right)v_{\ell} 
\label{acoplada1}
\end{eqnarray}
which can be easily decoupled, becoming second order differential equations.

The uncoupled equation for $u_{\ell}$, after making $x=\cos\theta$, is
\begin{eqnarray}
\left[\partial_x (1-x^2) \partial_x - \frac{(\ell -Ax)^2 + \frac{1}{4} + A - \ell x}{1-x^2}\right] u_{\ell}\nonumber \\ 
= \left[- \frac{1}{v_F^2\hbar^2}\left(E+\ell \hbar \Omega +\frac{s \hbar \Omega}{2}\right)^2+\frac{1}{4}\right]u_{\ell}. 
\label{eq}
\end{eqnarray}
The asymptotic behavior of this equation suggests  the following $ansatz$
\begin{equation}
u_j = (1-x)^\mu (1+x)^\nu \overline{u}_j (x),
\label{asymp}
\end{equation}
where
\begin{equation}
\mu = \frac{1}{2}\left| \ell - A - \frac{1}{2} \right| , \nu = \frac{1}{2}\left| \ell + A + \frac{1}{2} \right| .
\end{equation}

Inserting (\ref{asymp}) in (\ref{eq}) results in the following equation for $\overline{u}_j (x)$
\begin{eqnarray}
(1-x^2)\partial^{2}_{x} \overline{u}_{\ell} + [2(\nu - \mu) - 2(\mu + \nu +1)x]\partial_x \overline{u}_{\ell} \nonumber \\ + \left(- 2 \mu \nu - \mu -\nu - \frac{1}{2}(\ell^2 - A^2 + \frac{1}{4} - A)  \right.\nonumber \\
 \left. + \frac{1}{v_F^2\hbar^2}\left(E+\ell \hbar \Omega +\frac{s \hbar \Omega}{2}\right)^2-\frac{1}{4}\right)\overline{u}_{\ell} = 0,
\label{hiper}
\end{eqnarray}
which can be rewritten as
\begin{equation}
z(1-z)\partial^{2}_{x}\overline{u}_{\ell} + [c-(a+b+1)z]\partial_x \overline{u}_{\ell} - ab\overline{u}_{\ell} =0,
\label{hiper1}
\end{equation}
where
\begin{eqnarray}
a&=&\frac{1}{2}+\mu +\nu -\frac{1}{2}\left(4\mu^2 (4+8\nu)\mu +4\nu^2 +4\nu +1+4\alpha \right)^{\frac{1}{2}}, \nonumber \\
b&=&\frac{1}{2}+\mu +\nu +\frac{1}{2}\left(4\mu^2 (4+8\nu)\mu +4\nu^2 +4\nu +1+4\alpha \right)^{\frac{1}{2}}, \nonumber \\
c&=&2\nu +1 ,\\
z&=&\frac{1}{2}+\frac{1}{2}x ,\nonumber
\end{eqnarray}
and 
\begin{eqnarray}
\alpha &=&- 2 \mu \nu - \mu -\nu - \frac{1}{2}(\ell^2 - A^2 + \frac{1}{4} - A) \nonumber \\
&&+ \frac{1}{v_F^2\hbar^2}\left(E+\ell \hbar \Omega +\frac{s \hbar \Omega}{2}\right)^2-\frac{1}{4} .
\end{eqnarray}

Equation (\ref{hiper1}) is clearly the hypergeometric equation whose solution is
\begin{eqnarray}
\overline{u}_j&=&C_1\;\; {}_2F_1(a,b;c;z) \nonumber \\
&&+C_2z^{1-c}{}_2F_1(a+1-c,b+1-c;2-c;z),
\end{eqnarray}
where $C_1$ and $C_2$ are normalization constants and ${}_2F_1$ is the hypergeometric function.

Physics requires that the wave function  be square integrable. So, the hypergeometric series has to be convergent. This happens for arbitrary $a$, $b$ and $c$ when $-1<z<1$, and for $c>a+b$ when $z=\pm1$. As $-1\leq x\leq 1$, we have that $0\leq z\leq 1$. Therefore, in order to have a convergent series, the condition $c>a+b$ has to be satisfied.

We have that
\begin{equation}
a+b=1+2\mu + 2\nu > c 
\end{equation}
because $\mu$ is a positive constant. So, we have a divergent solution.

This problem is solved when we choose $a=-n$, where $n$ is an integer, which guarantees that we have a polynomial solution of degree $n$. From this condition, we obtain the energy spectrum as
\begin{eqnarray}
E^{(u)}_{n,\ell ,s}=\hbar v_F\biggl[\biggl(n&+&\mu +\nu +\frac{1}{2}\biggr)^2 \nonumber \\ 
&-&A-A^2\biggr]^{1/2}-\left(\ell +\frac{s}{2}\right)\hbar \Omega \; .
\label{spectrum1}
\end{eqnarray}
When $\Omega =0$ in (\ref{spectrum1}), one gets the energy spectrum obtained in \cite{osipov} for static fullerene.

Rotation adds  two terms to the energy spectrum. The first one, the energy shift $-\ell \hbar \Omega$, was already obtained in \cite{shen} from the Aharonov-Carmi phase\cite{arahonov.carmi, arahonov.carmi1}. The second is the spin-rotation coupling, analogous to the Zeeman coupling between magnetic field and spin.

Analogous to what as was done to $u_{\ell}$, one can uncouple Eqs.(\ref{acoplada}) and (\ref{acoplada1}) for $v_{\ell}$ and use the $ansatz$
\begin{equation}
v_{\ell} = (1-x)^\gamma (1+x)^\delta \overline{v}_{\ell} (x),
\end{equation}
where
\begin{equation}
\gamma = \frac{1}{2}\left| \ell - A + \frac{1}{2} \right| , \delta = \frac{1}{2}\left| \ell + A - \frac{1}{2} \right| \; ,
\end{equation}
to get another energy spectrum
\begin{eqnarray}
E^{(v)}_{n,\ell ,s}=\hbar v_F\biggl[\biggl(n&+&\gamma +\delta +\frac{1}{2}\biggr)^2 \nonumber \\
&-&A-A^2\biggr]^{1/2}-\left(\ell + \frac{s}{2}\right)\hbar \Omega .
\label{spectrum2}
\end{eqnarray}

The spectra (\ref{spectrum1}) and (\ref{spectrum2}) have to be the same, {\it i.e.}, the condition $E^{(u)}_{n,\ell}=E^{(v)}_{n,\ell}$ must be satisfied. There are two possible cases where this happens. The first one is when $\mu+\nu=\gamma+\delta$. In this case the possible values of $\ell$ are $|\ell|\geq ||A|+1/2|$. The energy spectrum becomes
\begin{eqnarray}
E_{n,\ell,s}=\hbar v_F\biggl[\biggl(n&+&|\ell| +\frac{1}{2}\biggr)^2   \nonumber \\
 &-& A-A^2\biggr]^{1/2}-\left(\ell +\frac{s}{2}\right)\hbar \Omega \label{spec}
\end{eqnarray}
and the eigenfunctions are
\begin{eqnarray}
u_{\ell}&=&(1-x)^\mu (1+x)^\nu [ C_1\;\; {}_2F_1(-n,b;c;z)  \nonumber \\
 &+&C_2z^{1-c}{}_2F_1(-n+1-c,b+1-c;2-c;z) ] \nonumber
\end{eqnarray}
and
\begin{eqnarray}
v_{\ell}=&&(1-x)^\gamma (1+x)^\delta [ C_1\;\; {}_2F_1(-n,b';c';z)  \nonumber \\
 &+&C_2z^{1-c'}{}_2F_1(-n+1-c',b'\nonumber \\
 &+&1-c';2-c';z) ] ,
\end{eqnarray}
where $c'$ and $b'$ have the same form as $c$ and $b$ with the exchange of $\mu$ by $\gamma$ and $\nu$ by $\delta$.

The second case is when 
\begin{equation}
E_{0,\ell , s}=-\left(\ell +\frac{s}{2}\right)\hbar \Omega, \label{nonzeromode}
\end{equation}
which makes  the right hand side of equations (\ref{acoplada}) and (\ref{acoplada1}) vanish. In this case, the possible values for $\ell$ and $n$ are determined by $|\ell|\leq ||A|-1/2|$ and $n=0$. Without rotation, this is a zero mode. Therefore,  rotation not only shifts the zero mode to $-\ell \hbar \Omega$ but also splits it into two states. The eigenfunctions for this case are given by
\begin{eqnarray}
u_{\ell}=0, \;\;\;\;\;v_{\ell}=C (1-x)^\gamma (1+x)^\delta .
\label{zero}
\end{eqnarray}
Since $A=\frac{3}{2}k+1=-1/2, 5/2$ and we must have $|\ell|\leq ||A|-1/2|$ then, when $A=-1/2$, $\ell =0$, and when $A=5/2$, $\ell =0,\pm 1, \pm 2$. Therefore the degeneracy of the states given by (\ref{nonzeromode})  would be $6$ but the Zeeman splitting reduces it to $3$.

The overall spectrum obtained in this continuum approach is given by Eqs. (\ref{spec}) and (\ref{nonzeromode}). Both equations indicate that the contribution of rotation  is a shift of $-\left(\ell +s_z\right)\hbar \Omega$ to the static molecule spectrum. We observe that  the integer part, $\ell \hbar \Omega$, of the shift has been predicted 
in reference \cite{shen} using an approximation based on the Aharonov-Carmi effect\cite{arahonov.carmi1,arahonov.carmi}. The non-integer part of the shift, which includes a splitting of levels analogous to  Zeeman's, is related to the geometric phase acquired by the electrons in rotating $C_{60}$ studied in \cite{shen2}. Both effects derive from the spin-rotation coupling.  

\section{Concluding Remarks}
In this paper, we report the results of the electronic structure of  rotating fullerene obtained using the continuum model based in an effective Dirac equation. The model   includes fictitious non-Abelian gauge field fluxes which simulate the pentagons of the $C_{60}$ molecule. There are two main differences between the energy spectrum here obtained and the static ones from previous studies. The first one  is a shift that comes from the coupling between the orbital angular momentum and rotation and which removes the zero modes. The second one is a Zeeman-like splitting of the levels due to the coupling between the real spin and rotation. This suggests the exciting possibility of observing Electron Spin Resonance without a magnetic field by tuning the rotation speed with a variation of temperature.  The occurrence of rapidly rotating molecules in solid fullerene is an experimental reality and therefore we expect that our work may contribute to a better understanding of the electronic properties of those fascinating solids. Moreover, a rotating system is of course accelerated and therefore, non-inertial. Non-inertial systems are undistinguishable from systems in the presence of a gravitational field, according to the Equivalence Principle of General Relativity. So, what in fact we are obtaining in the present work is a gravitational effect in  quantum mechanics which may be observed with conventional experimental techniques. This problem becomes even more interesting if one considers  an accelerated rotation and also the presence of a magnetic field, which we expect to address in  forthcoming publications.

{\bf Acknowledgements}: This work was partially supported by  CNPq, CNPQ-MICINN binational, INCTFCx, CAPES-WEISSMANN binational and CAPES-NANOBIOTEC. We are also grateful to Maria Vozmediano for the hospitality and illuminating discussions during the stay of JRFL and FM at Instituto de Ciencias de los Materiales de Madrid, where this work began.

%

\end{document}